\documentclass[review, onecolumn,english,prl,preprint,superscriptaddress]{revtex4}
\usepackage[LGR,T1]{fontenc}
\usepackage[latin1, utf8]{inputenc}
\usepackage{textgreek}
\usepackage{color}
\usepackage{multirow}
\usepackage{makecell}
 \usepackage{color,soul}
\setcitestyle{super}
\usepackage{amsmath}
\usepackage{amssymb}
\usepackage{graphicx}
\usepackage{textcomp}
\usepackage{array}
\usepackage{blindtext}
\newcolumntype{L}{>{\centering\arraybackslash}m{2cm}}
\makeatletter
\newcommand\emailx[1]{%
\move@AF%
\def\@affil{{\normalfont\,#1\strut}{}}%
}%

\makeatletter

\ProvideTextCommand{\~}{LGR}[1]{\char126#1}

\@ifundefined{textcolor}{}
{%
 \definecolor{BLACK}{gray}{0}
 \definecolor{WHITE}{gray}{1}
 \definecolor{RED}{rgb}{1,0,0}
 \definecolor{GREEN}{rgb}{0,1,0}
 \definecolor{BLUE}{rgb}{0,0,1}
 \definecolor{CYAN}{cmyk}{1,0,0,0}
 \definecolor{MAGENTA}{cmyk}{0,1,0,0}
 \definecolor{YELLOW}{cmyk}{0,0,1,0}
}

\usepackage[english]{babel}

\def\url#1{}
\addto\captionsenglish{%
}

\usepackage[none]{hyphenat}
\usepackage{multirow}

\usepackage{babel}

\makeatother

\begin{document}

\title{2D Spintronics for Neuromorphic Computing with Scalability and Energy Efficiency}

\author{Douglas Z. Plummer}
\affiliation{Department of Electrical and Computer Engineering, University of Florida, Gainesville, Florida 32611, USA.}
\author{Emily D'Alessandro}
\affiliation{Department of Electrical and Computer Engineering, University of Florida, Gainesville, Florida 32611, USA.}
\author{Aidan Burrowes}
\author{Joshua Fleischer}
\author{Alexander M. Heard}
\affiliation{Department of Electrical and Computer Engineering, University of Florida, Gainesville, Florida 32611, USA.}
\author{Yingying Wu\footnote{Corresponding author: yingyingwu@ufl.edu}}
\affiliation{Department of Electrical and Computer Engineering, University of
Florida, Gainesville, Florida 32611,
USA.}

\begin{abstract}
The demand for computing power has been growing exponentially with the rise of artificial intelligence (AI), machine learning, and the Internet of Things (IoT). This growth requires unconventional computing primitives that prioritize energy efficiency, while also addressing the critical need for scalability. Neuromorphic computing, inspired by the biological brain, offers a transformative paradigm for addressing these challenges. This review paper provides an overview of advancements in 2D spintronics and device architectures designed for neuromorphic applications, with a focus on techniques such as spin-orbit torque, magnetic tunnel junctions, and skyrmions. Emerging van der Waals materials like CrI$_3$, Fe$_3$GaTe$_2$, and graphene-based heterostructures have demonstrated unparalleled potential for integrating memory and logic at the atomic scale. This work highlights technologies with ultra-low energy consumption (0.14 fJ/operation), high switching speeds (sub-nanosecond), and scalability to sub-20 nm footprints. It covers key material innovations and the role of spintronic effects in enabling compact, energy-efficient neuromorphic systems, providing a foundation for advancing scalable, next-generation computing architectures.
\end{abstract}

\maketitle

\newpage

\section{Introduction}

\begin{figure}[h]
\vspace{-15pt}
    \centering
    \includegraphics[width=0.8\linewidth]{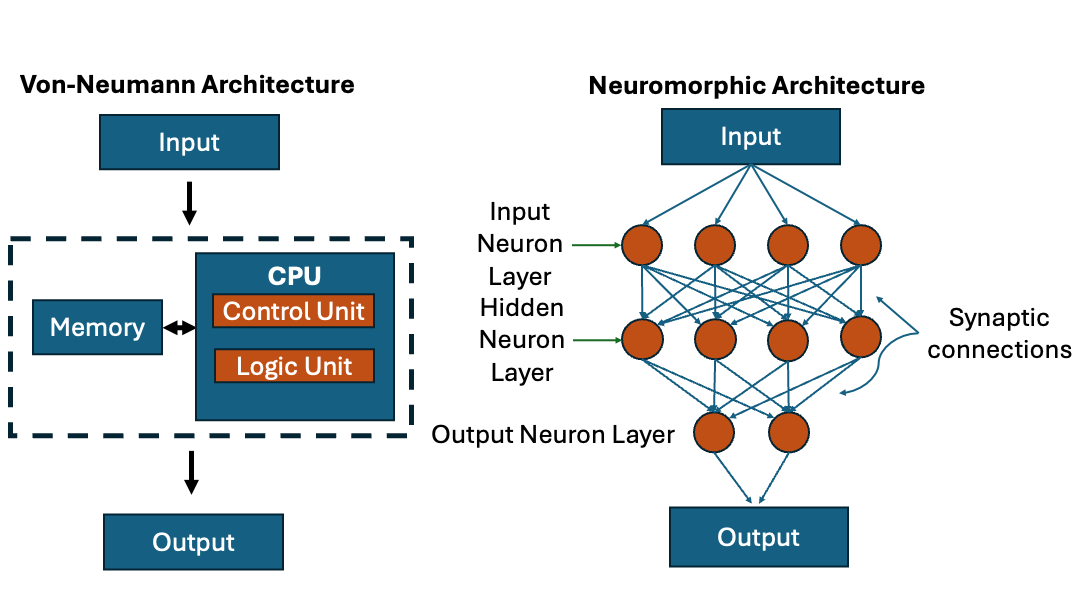}
    \vspace{-30pt}
    \caption{Comparison between the traditional von Neumann architecture and neuromorphic computing architecture. The von Neumann model (left), made up of a separate memory and processing unit (CPU), where data shuttling between memory and computation causes a bottleneck, limiting speed and energy efficiency. The neuromorphic architecture (right) mimics biological neural networks, integrating computation and memory within synaptic connections. The interconnected neuron layers (input, hidden, and output) enable parallel processing and event driven computation, significantly improving power efficiency and adaptability. Computing paradigms are enabled by 2D spintronic neuromorphic systems, which promise further enhancements in scalability and energy efficiency.}
    \label{fig:Von}
    \vspace{-15pt}
\end{figure}

Neuromorphic computing represents a burgeoning approach to computational architecture and device synthesis. Aimed at capturing the human brain's efficiency, adaptability, and massive parallelism, neuromorphic computing devices differ from the conventional von Neumann architecture by integrating memory and computation in a unified framework (Fig. \ref{fig:Von}). In spintronic neuromorphic systems, critical material and device properties include low magnetic damping for energy efficiency, high spin polarization for improved signal fidelity, and stable magnetic anisotropy to maintain data integrity. Magnetic materials play a crucial role due to their fast switching speeds and non-volatility, which facilitate synaptic behavior and memory retention. As CMOS technology approaches the limits of Moore's Law, current CMOS-based systems face substantial challenges in scalability and power consumption. Further miniaturization is hindered by increased leakage currents and heat dissipation, reducing energy efficiency. Moreover, the von Neumann bottleneck, which separates computation and memory, imposes fundamental limits on processing speed and increases latency and power consumption. Neuromorphic computing addresses these issues by enabling data-intensive tasks like pattern recognition and sensory processing with significantly lower energy requirements. Among various material systems explored for neuromorphic devices, 2D materials, van der Waals (vdW) heterostructures, and spintronic phenomena have emerged as key enablers, offering unique advantages in scalability, energy efficiency, and novel device functionalities.

2D materials\cite{chen2016probing} such as semimetal graphene\cite{wu2020large}, semiconducting WSe$_2$\cite{wu2019induced}, superconducting NbSe$_2$\cite{han2018investigation} and magnetic CrI$_3$\cite{Gong_2017} have atomically thin structures with unique properties, including high electron mobility (1,000,000 cm$^2$/V$\cdot$s), low switching current, and layer-dependent direct/indirect band gaps. These features enable 2D materials to serve as active layers in devices that mimic the synaptic weighting seen in neural networks. For instance, MoS$_2$-based memristors have achieved ultrafast switching speeds ($<$40 ns) and low power consumption ($\sim$18 fJ per pulse), making them suitable for high-density neuromorphic networks\cite{Cao2021, Wang2020a}. Meanwhile, h-BN/graphene heterostructures provide excellent retention and endurance, critical for stable neuromorphic operation\cite{Wang2020a}.

The integration of 2D magnetic materials - 2D magnet, into vdW heterostructures allows for the creation of devices with enhanced functionality and scalability. These heterostructures exhibit dangling-bond-free interfaces, which minimize scattering and enable efficient charge and spin transport, crucial for spintronic applications\cite{Zhang_2022}. For example, graphene/antiferromagnet heterostructures can help induce exchange splitting into graphene and enable long-range spin transport\cite{wu2020large}.

Similarly, spintronics leverages the spin of electrons rather than their charge, a key factor in achieving low-power synapse-based computation. Spintronic phenomena, such as magnetic tunnel junctions (MTJs), low current density-driven Skyrmion control, show great promise for neuromorphic computing. MTJs devices utilize tunneling magnetoresistance (TMR) to store and manipulate information. The integration of 2D materials like graphene electrodes reduces power consumption and enhances scalability\cite{Yao_2023}. Magnetic skyrmions, nanoscale spin vortices, can be manipulated experimentally to emulate synaptic plasticity while maintaining ultra-low power consumption (as low as 0.14 fJ per operation) and excellent scalability\cite{Chen_2018}. Materials such as Fe$_3$GaTe$_2$ and MnSi have demonstrated room-temperature skyrmion stability, opening avenues for spin-based neuromorphic computing in practical applications.
2D materials (such as graphene, TMDs, and MXenes) excel in scalability, high-speed operation, energy efficiency, and flexibility. These materials are particularly advantageous for low-power, high-speed operations and their integration into compact, flexible neuromorphic devices. In addition to sharing these benefits, 2D spintronics using 2D magnets offer unique advantages in memory storage, non-volatility, and stability. This makes them particularly well-suited for long-term data retention, energy-efficient memory operations, and high parallelism-key features for advanced neuromorphic computing systems.

This review paper summarizes recent progress in the use of spintronics for neuromorphic computing, with a particular focus on the outlook and future directions for integrating 2D spintronic materials into neuromorphic hardware. It covers the fundamentals of neuromorphic computing, an overview of 2D spintronic materials, and existing spintronic devices and architectures, such as those based on MTJs and magnetic skyrmions. Additionally, it discusses the scalability and energy efficiency of current spintronic neuromorphic systems and explores potential applications of 2D spintronics in neuromorphic computing.

\section{Fundamentals of Neuromorphic Computing}
Neuromorphic computing relies on two key principles\cite{ma2025van,rehman20252d,hadke2025two}: spiking neural networks (SNNs) and synaptic plasticity. SNNs process discrete events, or spikes, instead of continuous signals, enabling asynchronous, event-driven information processing. This mechanism replicates the biological firing of neurons, where a spike occurs only when a threshold potential is reached. In SNNs, an output spike is generated only when the membrane potential exceeds a defined threshold. This event-driven mechanism allows for synchronous processing, where consumption occur only when necessary, reducing energy consumption\cite{li2021fast}. As a result, SNNs have much lower idle power consumption compared to traditional artificial neural networks (ANNs). For example, second-generation networks like deep learning networks (DLNs) process real-valued signals continuously, whereas SNNs process binary events, achieving energy savings of up to 20 times in certain cases\cite{Yang2021, Zhou2021}. Synaptic plasticity refers to the dynamic adjustment of synaptic weights based on temporal activity, a key principle to enable adaptive learning in neuromorphic systems. Synaptic weights determine the strength of the connection between neurons, influencing how input spikes affect the receiving neuron's membrane potential. Artificial implementations utilize materials with tunable resistance and magnetization, particularly those based on Spin-orbit torque (SOT)-magnetoresistive random-access memory (MRAM), allowing energy-efficient weight updates with costs as low as 10 fJ per state change. This offers a significant advantage over charge-based memories in neuromorphic applications\cite{sosa2025simulating,CHEN2023193, Zhou2021}. Weight updates in synaptic plasticity are typically governed by biological learning rules, such as spike-timing-dependent plasticity (STDP). In STDP, the timing relationship between pre-synaptic and post-synaptic spikes determines whether synaptic weights are strengthened or weakened. This temporally sensitive learning rules enables neuromorphic systems to adopt to changing input patterns and support efficient, continuous learning.

2D spintronics refers to the study and application of spin-based electronic phenomena in 2D materials including atomically thin vdW layers\cite{lin2019two}. It offers unique advantages for neuromorphic computing. SOT devices demonstrate switching energy as low as 10 fJ per event. Additionally, domain wall motion in magnetic nanowires enables highly energy-efficient state manipulation, achieving motion velocities of $\sim$5700 m/s under 1 ns pulses\cite{Yang2021, Zhou2021}. The ability to retain information without continuous power consumption is intrinsic to spintronic devices. MTJs, for example, exhibit nonvolatile data retention that exceeds 10 years at room temperature. This feature reduces energy consumption and aligns with the memory functionality of biological synapses. The 2D nature of spintronic materials enables ultra-thin device layers, supporting high-density integration. For instance, MTJs and skyrmionic devices achieve storage surpassing 1 \(\text{Tb/in}^2\), making them suitable for scalable neuromorphic hardware\cite{CHEN2023193, Yang2021,zhong2024integrating,zhang2d2024}. These nanosecond-scale switching times make spintronics significant for real-time computation. Real-time computation refers to the ability of a system to process and respond to inputs within a time frame that meets the requirements of the task, typically without perceptible delay. In the context of 2D spintronics, real-time computation is achieved through the rapid manipulation of spin states, enabling fast data processing and immediate responses to incoming information. That is why nanosecond-scale switching time of spintronic system is preferred for real-time computation. This rapid switching allows these devices to perform logic operations and memory updates at extremely high speeds, aligning with the real-time demands of neuromorphic hardware. The advantage of using 2D spintronics for high density devices arises from the following key aspects\cite{zhang2d2024}: (1) monolayer high-quality 2D materials that are free from dangling bonds, unlike thin films grown via sputtering or molecular epitaxy growth, which often result in non-uniform isolating islands; (2) seemless heterogeneous integration with most interface, eliminating lattice mismatch issues that typically occur in as-grown layers; and (3) exceptional scalability, as 2D materials can be fabricated or exfoliated at the the nanometer scale and compatible with advanced fabrication techniques. This makes them ideal for producing ultra-compact devices while maintaining energy efficiency and speed.

Additionally, their inherent stability ensures consistent performance, crucial for complex neuromorphic tasks like pattern recognition and adaptive control. In hardware-accelerated SNNs, SOT-MRAM synapses have been integrated into crossbar arrays\cite{verma2023neuromorphic}, where the conductance of each MTJ emulates synaptic weight adjustments. Recent experimental demonstrations\cite{sosa2025simulating,verma2024multi} have shown SOT-MRAM arrays achieving high classification accuracy in pattern recognition tasks while consuming up to 85\% less power than static random-access memory (SRAM)-based implementations. Furthermore, reservoir computing architectures, which rely on the inherent memory properties of spintronic devices, have leveraged SOT-driven magnetic textures (such as domain walls and skyrmions) to process temporal data with minimal energy overhead. In one reported experiment, SOT-MRAM-based reservoirs successfully implemented real-time gesture recognition tasks at sub-mW power levels, highlighting their viability for energy-efficient edge AI applications\cite{CHEN2023193}. By aligning the biological principles of SNNs and synaptic plasticity with the energy efficiency, non-volatility, and scalability of spintronic/MRAM materials, neuromorphic computing stands to overcome the limitations of von Neumann architectures. 

\section{Overview of {2D} Spintronic Materials}
\begin{figure}[h]
    \centering
    \includegraphics[width=1\linewidth]{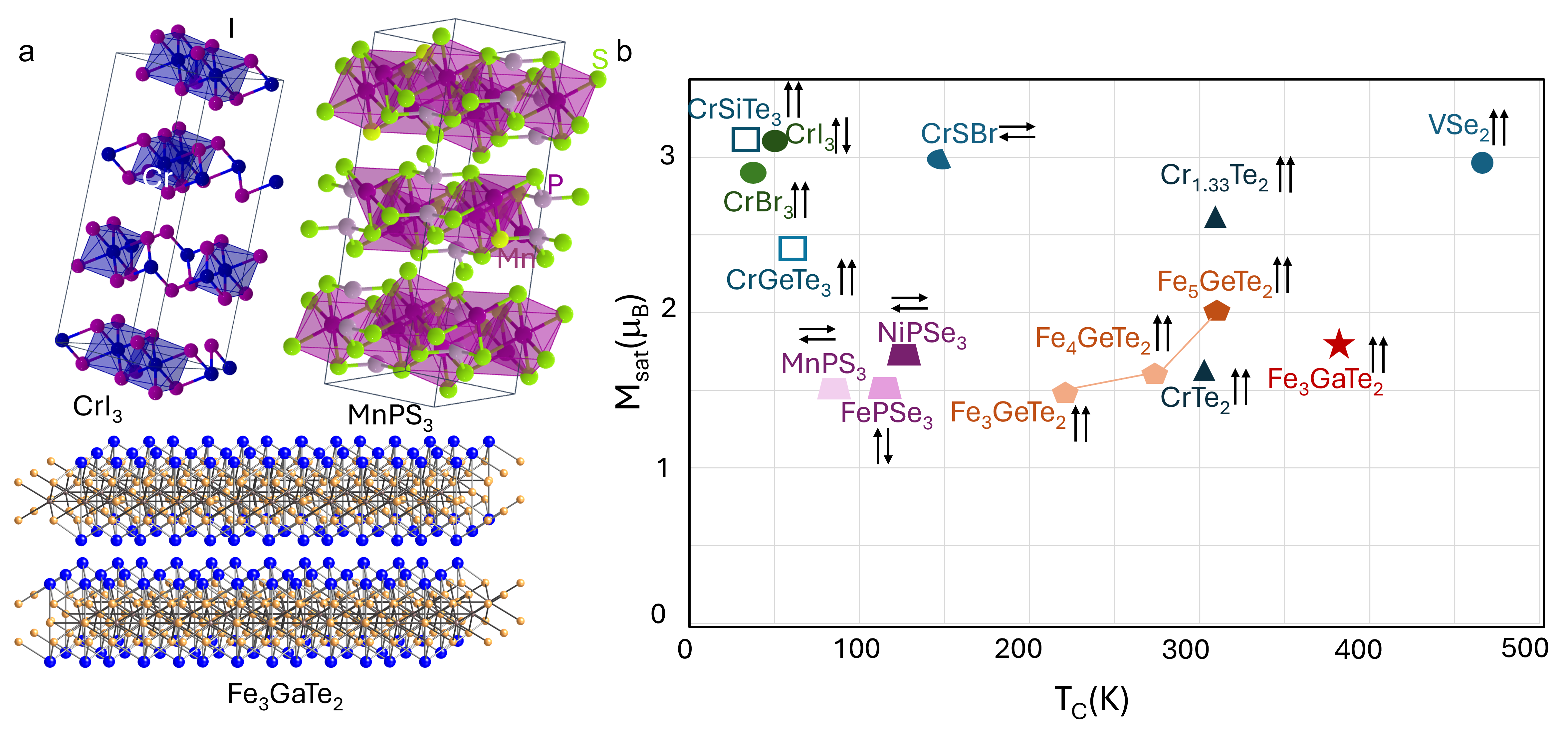}
    \caption{2D magnetic materials. (a) Crystal structure of CrI$_3$, Fe$_3$GaTe$_2$ and MnPS$_3$. The above two structures are adopted from Materials Project. (b) Critical temperature versus saturated magnetization in 2D antiferromagnets and ferromagnets. }
    \label{fig:material}
\end{figure}

2D magnets, such as CrI$_3$, Fe$_3$GaTe$_2$, and MnPS$_3$, are promising for scalable, energy-efficient neuromorphic architectures. Their atomic-thin magnetic ordering allows for ultra-dense integration of spintronic devices with low power consumption, enabling non-volatile, tunable synaptic elements and neuronal functionalities\cite{Wu2021n}. CrI$_3$ features intrinsic ferromagnetism and strong out-of-plane magnetization below 68 K, with tunable magnetic states in bilayers, making it ideal for nanoscale devices (Fig. \ref{fig:material}). It also exhibits second-order topological insulating behavior, offering stable platforms for spintronic innovations and information storage\cite{LIU2024594}. Fe$_3$GaTe$_2$ overcomes a key limitation by showing intrinsic ferromagnetic behavior at room temperature (350-380 K), with high magnetic moment and perpendicular magnetic anisotropy \cite{wu2024room,zhang2d2024}, making it ideal for spintronic devices like MTJs. Its strong spin-orbit coupling (SOC) enables efficient room-temperature magnetization switching, enhancing low-power electronics\cite{Zhang2022}. MnPS$_3$, a 2D antiferromagnet, offers ultra-low-power synaptic emulation with its stable in-plane antiferromagnetic ordering, fast switching speeds (20 ns), and minimal power dissipation. It also supports multi-bit synaptic weight encoding and offers potential for spin liquids, opening doors for quantum-inspired neuromorphic computing\cite{Burch2018}. Among the discovered 2D magnet (Fig. \ref{fig:material}b), VSe$_2$ has the highest critical temperature of over 400 K\cite{bonilla2018strong}. However, its room-temperature ferromagnetism has been difficult to reproduce in other studies. In contrast, 2D antiferromagnets generally have much lower critical temperatures, well below room temperature. Therefore, there is a need for the discovery of high-temperature antiferromagnets, like combining machine learning for materials discovery and characterization\cite{leger2024machine}. These materials' atomic thickness enables high-density integration, and their low energy requirements for magnetization switching, tunability, and scalability make them critical for advancing spintronic technologies\cite{Burch2018, Zhang2019}.

SOC and topological properties in 2D materials\cite{wang2020topological} have emerged as vital enablers for energy-efficient and scalable neuromorphic computing. One promising avenue involves topological insulators (TIs) and quantum anomalous Hall insulators (QAHIs), which support large charge-to-spin interconversion (CSI) and exhibit large spin Hall conductivities. Such traits are especially attractive in applications requiring rapid, ultra-low-power synaptic updates\cite{Vil2020LowsymmetryTM, Yasuda2016CurrentNonlinearHE}. For instance, TIs have demonstrated high spin Hall conductivity favorable for spin logic, while QAHIs offer large CSI, a property pivotal for fast, energy-saving manipulation of synaptic weights in artificial neural networks\cite{Vil2020LowsymmetryTM}. Beyond TIs, recent investigations into WTe$_2$ and Bi$_2$Se$_3$ highlight the potential of 2D materials with strong SOC for neuromorphic devices. WTe$_2$, for example, features high spin-torque efficiency, enabling field-free magnetization switching at moderate currents\cite{Binda2021a}. Meanwhile, Bi$_2$Se$_3$ has garnered attention for its large CSI, which is critical for high-speed spin injection and readout\cite{Wang2015TopologicalSS}. Notably, devices employing WTe$_2$ have demonstrated room-temperature operation with damping-like and field-like spin-torque efficiencies of 0.12 and 0.18, respectively\cite{Binda2021a}, underscoring their feasibility for practical neuromorphic systems.

Continued progress in material engineering has led to synaptic devices achieving sub-nanosecond switching and per-bit energies under 10 fJ, thus rivaling the exceptionally high throughput and low power consumption of biological synapses\cite{Singh2024, Merkel2023SynapticSA}. Concurrently, exotic spintronic designs, such as a circular bilayer skyrmion system, have demonstrated symmetric and linear long-term potentiation (LTP) and long-term depression (LTD) processes with energies as low as 0.87 fJ\cite{Gupta2024ACC}. These advances highlight how refined control over SOC can stabilize complex spin textures (e.g., skyrmions) to mimic synaptic behavior with minimal energy overhead. Leveraging heavy-metal/ferromagnet interfaces\cite{wu2020neel}, such as WTe$_2$/Fe$_3$GeTe$_2$ and WTe$_2$/Fe$_3$GaTe$_2$, prove effective for controlling spin waves and spin-torque efficiencies\cite{marfoua2024highly,shin2022spin,wang2023room}. Such material systems excel in tunability and integrability, offering a key stepping stone toward ultra-dense, low-power neuromorphic hardware.

\section{Spintronic Device Architectures for Neuromorphic Computing}

The integration of spintronic devices into computing systems relies on harnessing the spin degree of freedom of electrons, in addition to their charge, to process and store information. This is achieved through key components like MTJs and spin-transfer torque (STT) or SOT mechanisms, which enable data manipulation with lower energy consumption compared to conventional charge-based electronics. Spintronic devices offer non-volatility, allowing data to be retained without continuous power, which enhances energy efficiency. Furthermore, they enable parallel processing and in-memory computation, reducing the need for data transfer between memory and processors - overcoming the von Neumann bottleneck. In terms of scalability, spintronic devices can be miniaturized to the nanoscale, making them suitable for high-density integration in advanced neuromorphic systems. Emerging 2D magnets further enhance scalability by enabling ultra-thin, flexible device architectures. However, challenges remain in maintaining consistent switching behavior, reducing error rates, and achieving high-speed operation at large scales. Successful integration of spintronics promises to improve computing efficiency by offering faster, more energy-efficient, and densely packed neuromorphic architectures, paving the way for future high-performance computing applications. 

\subsection{MTJs and Spin Valves}
The key memory functionality in novel spintronic devices relies on utilizing spin degrees of freedom to store and transmit data, a process increasingly enabled by MTJs. An MTJ consists of a simple material configuration with a tunneling dielectric layer sandwiched between two ferromagnetic layers. Magnesium oxide (MgO) is commonly used as the tunneling dielectric due to its high TMR ratio\cite{Mathon2006, Jia2011}. In 2D case, h-BN is often used as the dielectric layers, with Fe$_3$GeTe$_2$ or Fe$_3$GaTe$_2$working as the ferromagnetic electrodes\cite{li2019spin,li2021large,zhang2020perfect,li2023tremendous}. In this setup, the magnetization of one ferromagnetic layer is kept constant, while the other is manipulated using a current-induced magnetic field or spin-polarized currents via spin transfer torque (STT) or SOTs\cite{Tehrani2003, CHEN2023193}. This manipulation causes device resistance to fluctuate, encoding data as bits (0 or 1). The size of these `bits' can be reduced by shrinking the MTJ and minimizing the size of access transistors\cite{Hu2011}. The scalability of MTJs extends further with the use of multi-state MTJ devices, enabling the representation of multiple bits within a single bit cell\cite{Cao2020a}. These multi-state devices not only increase density but also offer neuromorphic computing functionality\cite{Indiveri2006}, as they can represent bit cell values in a weighted fashion, similar to the biological synaptic weighting mechanism.

\begin{figure}
    \centering
    \includegraphics[width=0.8\linewidth]{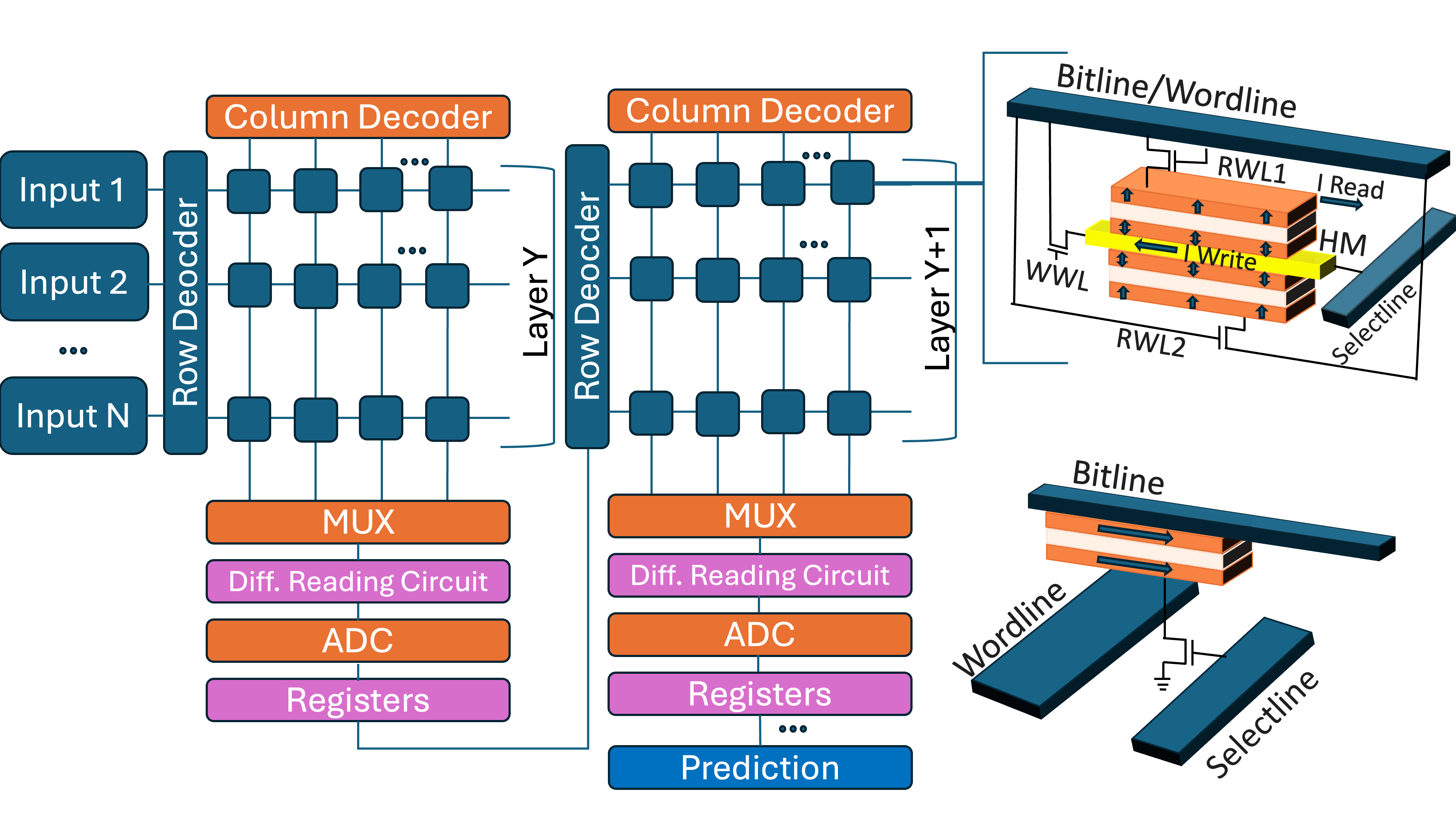}
    \caption{Block-level circuit implementation of a generalized multi-level MRAM synaptic device. The schematic illustrates two distinct MTJ-based transistor configurations: a series-configured dual-level cell (SOT-sDLC), which can also be adapted into a parallel or sVCMA STT/SOT-compatible arrangement, and a circuit diagram of an STT-MRAM bit cell employing a 1-transistor-1-MTJ (1T1J) single-cell configuration, traditionally used for magnetic memory applications.}
    \label{fig:MTJ Circuit Level Neuro}
\end{figure}

A key feature of multi-state MTJ devices is the use of domain wall motion, the transition region between two oppositely magnetized domains, to enable stable memristive or synaptic functionality, using 2D magnets\cite{younis2025magnetoresistance}. Many of these devices are driven by two-terminal STT mechanisms. For example, researchers\cite{Lim2004} studied domain wall motion in an in-plane magnetization configuration, finding that current pulses as short as 0.41 ns could displace the domain wall by up to 300 nm in width and 30 $\mu$m across. Other studies explore various device temperatures and geometries to attenuate domain walls theoretically\cite{ALBAHRI2022168611, Ababei2021}. The challenge remains on control of domain wall motion in a way that enhances the stability of memristive MTJ multi-state devices. Another work\cite{Lequeux2016} demonstrated multi-state stability in a resistive switching configuration driven by STT in a perpendicularly polarized magnetic strip. The current controlled both the direction and displacement of the domain wall, allowing for manipulation of resistance as a function of current.

More recently, three-terminal SOT-driven multi-state i-MTJ devices have emerged, achieving high switching speeds (pulses shorter than 400 ps), robust endurance, and low read interference\cite{Cubukcu2018, Chen2017}. These configurations can clearly resolve up to four resistance states without the need for pinning steps, with key materials including Pt/CoFeB/MgO. However, there is a notable limitation in scalability, as micromagnetic simulations suggest that multiple resistance states may not be achievable when domain walls shrink below a certain width. This is due to the reliance on metastable multi-domain states, which many smaller devices fail to exhibit, instead showing mono-domain behavior. Nevertheless, some simulation studies suggest that even mono-domain devices could potentially support multi-state switching beyond binary\cite{Li2021}.

Although these simulation experiments offer a promising outlook, the focus of neuromorphic MTJ efforts is on adapting the previously discussed domain wall memory devices to function as neurons\cite{Lone2022, Liu2021a, Lone_2023}. The most common manifestation of these neuron configurations is in the form of "step neurons", which receive inputs from other neurons and output a varying signal based on all previous inputs. This behavior has been simulated using an entirely MTJ-based device, where the domain wall position determines the magnetoresistance\cite{Sengupta2016}. Similarly, synapses can be constructed by leveraging the same MTJ-based domain wall device to exhibit multiple resistance states based on the domain wall position. The interplay between neurons and synapses in spintronic devices has led to significant interest in developing algorithms, with SNNs and probabilistic computing being of particular focus.

In addition, hybrid MTJ-CMOS circuits have shown strong potential for implementing neuromorphic crossbar arrays, with MTJs handling memory storage and CMOS elements managing peripheral control\cite{Marrows2024}. These systems achieve exceptional power efficiency, with multiply-accumulate (MAC) operations consuming as little as 20 fJ per operation-orders of magnitude lower than traditional CMOS-only designs. They also demonstrate high accuracy; for instance, SOT oscillators utilizing MTJs have achieved speech recognition rates of around 99.6\% on the NIST TI-46 data corpus, matching the performance of state-of-the-art neural networks. Fig. \ref{fig:MTJ Circuit Level Neuro} shows an example of block-level circuit implementation of multi-level MRAM synaptic device, where 1T1J configuration is adopted.

\subsection{Skyrmion-Based Devices}
Magnetic skyrmions typically arise in non-centrosymmetric materials exhibiting strong SOC and Dzyaloshinskii-Moriya interaction (DMI)\cite{Heinze2011,wu2024magnetic}, as well as in specially engineered multilayer stacks\cite{Fert2017,wu2020neel,zhang2d2024,zhong2024integrating,wu2022van}. Initially observed in bulk chiral magnets such as MnSi\cite{Mu_hlbauer_2009} and FeGe \cite{Yu2011} at cryogenic temperatures, recent advances have enabled the stabilization of skyrmions at or about room temperature for leveraging engineered interfacial DMI\cite{Woo2016, Jiang_2016}, strain\cite{Tokunaga_2015}, and thickness \cite{Caretta2018} in 2D or quasi-2D ferromagnets. Furthermore, recent work demonstrates that skyrmions can be artificially introduced in 2D ferromagnets like Cr$_2$Ge$_2$Te$_2$ and Fe$_3$GeTe$_2$, using magnetic force microscopy (MFM)\cite{Mi2024RealSpaceTO, wu2022van}.

Consequently, leveraging the advantages of 2D or quasi-2D architectures, skyrmion-based devices have shown great potential for energy-efficient neuromorphic computing applications\cite{Lone2022SkyrmionMagneticTJ, Yu2020VoltagecontrolledSN,Bindal2022AntiferromagneticSB}. The spintronic devices based on skyrmions promise increased density and energy-efficient data storage as a result of their small nanometric size and topological protection.\,Skyrmion-based devices can emulate neuronal dynamics by exploiting skyrmion motion under thermal or magnetic anisotropy gradients. For example, an antiferromagnetic skyrmion-based neuron device\cite{bindal2022antiferromagnetic} utilizes thermal or perpendicular magnetic anisotropy gradients within a nanotrack to drive the skyrmions and achieve integrate-and-fire (LIF) behavior. The device has been shown to have an energy dissipation of 4.32 fJ per LIF operation.
\begin{figure}
    \centering
    \includegraphics[width=0.9\linewidth]{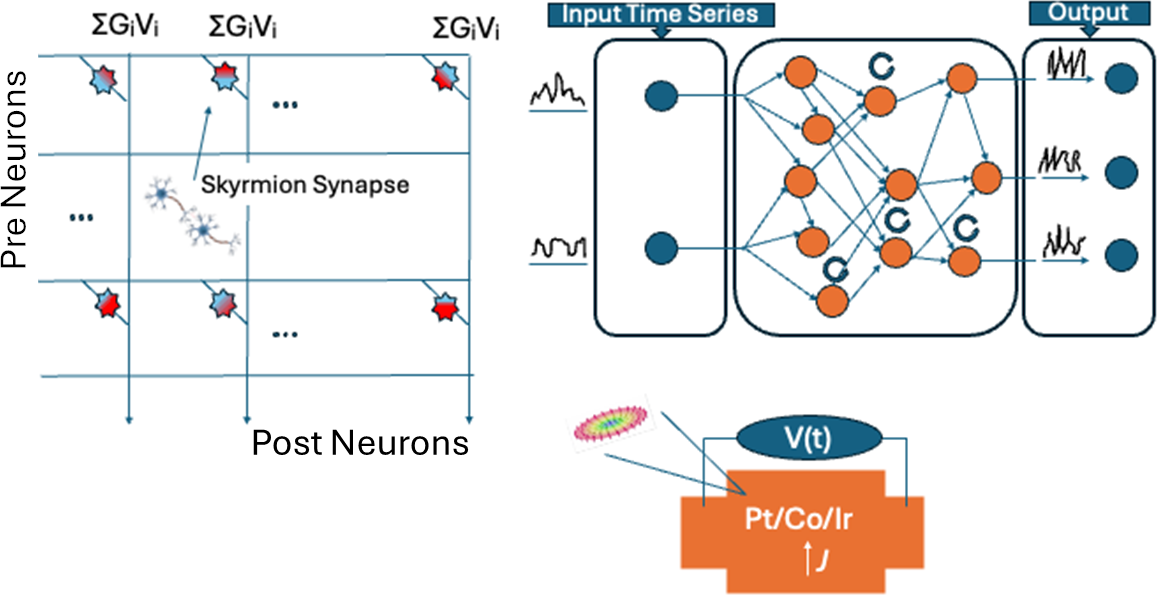}
    \vspace{-10pt}
    \caption{Two distinct skyrmion-based architectures designed for neuromorphic computing. The top left for schematic of a circuit diagram incorporating skyrmion-based synapses, where the conductance values at each level follow a normal distribution, with defined average and deviation whenever synaptic weights are updated. The right one with a schematic representation of a conventional reservoir computing model. Additionally, the device figure includes a diagram depicting a Hall bar device alongside a magnetic skyrmion, highlighting its structure and functional role within the system.}
    \label{fig:Skyrmion based Neurmorhic}
\end{figure}
Skyrmions in 2D materials can also support multi-level resistance states, analogous to synaptic weights\cite{Lone2023ControllingTS}. The device exhibits discrete topological resistance states controlled by SOT with high linearity and uniformity, allowing the implementation of the hardware implementation of weight quantization in a quantized convolutional neural network (QCNN)\cite{Lone2023ControllingTS}. The device has been shown to achieve a recognition accuracy of around 87\% on the CIFAR-10 dataset\cite{Lone2023ControllingTS}.
Another intriguing pathway involves voltage-controlled skyrmions in quasi-2D synthetic antiferromagnets (SAF) with a weak anisotopy energy\cite{Yu2019VoltagecontrolledSA, Yu2020VoltagecontrolledSN,wu2024room}. By applying a weak electric field to a heterostructure, interlayer antiferromagnetic coupling can be tuned, giving rise to a continuous transition between a large skyrmion bubble and a small skyrmion. Thus, this induces a variation in the resistance of an MTJ that can mimic the potentiation and depression of a synapse, as well as the leaky LIF function of a neuron at a very low energy consumption (approximately 0.3 fJ per operation)\cite{Yu2019VoltagecontrolledSA, Yu2020VoltagecontrolledSN}.

Furthermore, skyrmion-based architectures have been explored in recent years (Fig. \ref{fig:Skyrmion based Neurmorhic}). For example, the skyrmion-based synapse device integrates skyrmions as dynamic resistance elements within a circuit, where conductance states follow a normal distribution during weight updates\cite{Gupta2024ACC}. This device exhibits highly symmetric and linear weight modulation during LTP and LTD operations, achieving ultra-low energy consumption of approximately 0.87 fJ per state update. Evaluated for pattern recognition using the MNIST handwritten dataset, it achieves an accuracy of 98.07\%\cite{Gupta2024ACC}.

Alternatively, the skyrmion-based reservoir computing model (Fig. \ref{fig:Skyrmion based Neurmorhic}) utilizes thermally activated diffusive skyrmion motion to implement a computational reservoir\cite{Raab2022BrownianRC}. Experimentally demonstrated for Boolean logic operations, such as XOR gates, this architecture exploits thermal and current-driven spin dynamics. Its robustness against device imperfections and scalability—enabled by linking multiple confined geometries or incorporating additional skyrmions suggests strong potential for energy-efficient reservoir computing.

We must admit there is a lack of using 2D spintronics for architecture-level implementation of neuromorphic hardware. There are some works at the architecture level reporting the use of 2D materials with their memristive behaviors without magnetic properties\cite{joksas2022memristive} or spintronics for neuromorphic computing\cite{zhou2021prospect}. To bridge this gap, future research should focus on exploring the potential of 2D spintronic materials that combine both memristive behaviors and magnetic properties, enabling more efficient and scalable architecture-level implementations of neuromorphic computing. 

\section{Scalability and Energy Efficiency}
The feasibility of neuromorphic computing as a future architecture depends on achieving ultra-low-power operation while maintaining scalability in device design. Spintronic devices, particularly those leveraging 2D materials and novel magnetic phenomena, show great potential in overcoming these challenges. Table \ref{table:device-comp} analyzes specific configurations and materials that highlight their relevance to scalable and energy-efficient neuromorphic computing.

\begin{table}[h!]
    \centering
    \caption{Property comparison between devices/architectures.} 
    \resizebox{\textwidth}{!}{ 
    \begin{tabular}{|c|c|c|c|c|c|c|c|c|}
    \hline
        \textbf{Device} & \textbf{Efficiency} & \textbf{Scalability} & \textbf{Temp. (K)} & \textbf{Switch Speed} & \textbf{Footprint} & \textbf{Leakage} & \textbf{Error Rate} & \textbf{Refs.} \\ \hline \hline
        MTJ & 10–100 fJ/bit & >1 Tb/in$^{2}$ & 300 & ~1–10 ns & 20–50 nm  & NA & 10$^{-5}$ & \cite{Ikeda2010,Yuasa_2007} \\ \hline
        SOT Devices & 10–20 fJ/bit & Sub 10 nm & 300 & <1 ns & <50 nm & NA & 10$^{-5}$ & \cite{Miron2011,Khang2020} \\ \hline
        Spin Valves & 10$^{-16}$J/event & 10 nm & 300 & <10 ns & ~50 nm & NA & 10$^{-4}$ & \cite{Ghising2023,Wang2023} \\ \hline
        Skyrmions & 0.3-1 fJ/event & Sub 100 nm & 300 & ~100 m/s & <100 nm & NA & 10$^{-3}$ & \cite{Fert2017,Jiang2015} \\ \hline
        vdW HS & 2.5 fJ/event & $\sim$10nm & 300 & $\sim$40 ns & <10 nm & NA & 10$^{-2}$ & \cite{Zhang_2022,Tang_2021,Kim2021} \\ \hline
    \end{tabular}}
    \label{table:device-comp}
\end{table}

MTJ exhibit moderate energy efficiency (10-100 fJ/bit) and are highly scalable (>1 Tb/in$^{2}$), with a temperature stability up to 300 K. The switching speed of MTJs ranges from 1-10 ns, and they feature relatively small footprints (20-50 nm). MTJs also have a low error rate of approximately 10$^{-5}$, making them a reliable option for neuromorphic applications. SOT Devices offer higher energy efficiency (10-20 fJ/bit) and excellent scalability, with dimensions as small as <10 nm. They are characterized by fast switching speeds (less than 1 ns) and compact footprints (<50 nm). SOT devices also maintain a low error rate of 10$^{-5}$, making them suitable for high-performance, energy-efficient systems. Spin valves stand out for their ultra-low energy consumption (10$^{-16}$ J/event) and scalability to 10 nm. However, they have a slower switching speed (<10 ns) and a larger footprint ($\sim$50 nm). The error rate for spin valves is somewhat higher compared to MTJs and SOT devices, at approximately 10$^{-4}$, which may limit their use in certain applications. Skyrmions, with their energy efficiency of 0.3-1 fJ/event, are highly scalable (down to <100 nm). They have a moderate switching speed ($\sim$100 m/s) and a small footprint (<100 nm). However, their error rate (10$^{-3}$) could pose challenges in practical implementation, especially for sensitive neuromorphic systems. This error rate could be from stability of nanoscale skyrmions, thermal fluctuations, edge effect and pining. VdW heterostructures (vdW HS) show good energy efficiency of 2.5 fJ cost per event\cite{sun2022programmable} and scalability ($\sim$10 nm). While their switching speed is slower ($\sim$40 ns) and error rate is higher (10$^{-2}$) comprared to skyrmion case, their small footprint (<10 nm) makes them appealing for highly compact devices. The increased error rate may result from the impurities and defects at the interface.

Overall, skyrmions and spin valves offer the best energy efficiency, while SOT devices and MTJs excel in scalability, speed, and low error rates. SOT devices in particular stand out for their fast switching speed and small footprint, making them ideal candidates for high-performance, energy-efficient neuromorphic computing applications.

\section{Applications in Neuromorphic Computing}
2D spintronic devices are revolutionizing neuromorphic systems, enabling advancements in real-time learning, edge computing, and AI accelerators. MTJs, for example, exhibit impressive endurance, exceeding 10$^{15}$ write cycles, and maintain non-volatile data retention for over a decade, making them well-suited for energy-efficient, long-lasting neuromorphic applications\cite{Grollier2019,Joksas2022}. MTJ-based associative memories have achieved remarkable energy savings, with an 89\% reduction in energy consumption and a 13.6-fold decrease in memory requirements compared to traditional content-addressable memories. Additionally, these designs improve processing efficiency by reducing clock cycles for search operations by a factor of 8.6 and offering five orders of magnitude improvement over conventional processor-based systems\cite{Grollier2019,Zabihi2019}.

When integrated into crossbar arrays, 2D spintronic devices further enhance neuromorphic capabilities. These arrays use the resistive states of spintronic elements to emulate synaptic weights, enabling efficient computation of weighted sums - a critical operation in neural networks. Spintronic crossbars have demonstrated synaptic weight resolutions of up to 6 bits and resistive switching currents as low as 10 $\mu$A, facilitating ultra-low-power operation. MAC operations in these arrays consume as little as 20 fJ, significantly reducing energy usage compared to traditional digital systems\cite{Shumilin2024,Joksas2022}.

Spintronic devices have also transformed processing-in-memory (PIM) architectures, particularly in computational random-access memory (CRAM) systems. By integrating MTJs directly into memory arrays, spintronic CRAM enables logic operations to be performed in situ, reducing data transfer energy by 70\% and improving latency by a factor of 5 compared to near-memory processing architectures. In addition, spintronic CRAM has demonstrated significant performance gains in tasks like convolution operations and neural inference, achieving energy savings of up to 85\% compared to CPU-based implementations, while maintaining near-perfect accuracy in tasks like digit recognition on the MNIST database\cite{Zabihi2019}.

In wearable and edge computing applications, the compactness, low power consumption, and scalability of spintronic devices offer transformative benefits. These devices enable real-time, localized computation under tight energy budgets. Spintronic devices with perpendicular magnetic anisotropy have demonstrated switching energies as low as 100 fJ and switching times under 2 ns, making them ideal for wearable devices and AI accelerators that require rapid, energy-efficient computation. For edge computing platforms, spintronic neuromorphic systems operate at less than 1 W of power, reducing dependence on cloud infrastructure and enabling latency-critical tasks\cite{Zabihi2019}.

The combination of high-speed operation, energy efficiency, and scalability makes 2D spintronic devices key enablers for the advancement of neuromorphic computing. Their ability to integrate memory and computation on a nanoscale addresses fundamental limitations of traditional systems, paving the way for advanced, low-power AI solutions in wearable, edge, and real-time applications. On one hand, spintronics will drive the advancement of machine learning operations\cite{sosa2025simulating}, while on the other hand, machine learning techniques can be leveraged to optimize fabrication processes and device testing\cite{leger2024machine}, creating a mutually beneficial feedback loop.

\section{Challenges and Future Outlooks}
Despite significant progress in 2D spintronic devices, several challenges remain for their integration into neuromorphic computing systems. One primary challenge is the synthesis of high-quality 2D materials with consistent magnetic properties. Current techniques such as chemical vapor deposition (CVD) and mechanical exfoliation often result in structural defects and limited scalability, which can hinder device performance, especially in applications requiring precise manipulation of spin states\cite{Kumari2021, Lin2022}. For instance, the scalability of heterostructures, which rely on integrating multiple 2D layers, remains a bottleneck due to interface defects and uncontrolled doping levels\cite{Hao2022}.

The stability of magnetic properties at room temperature is another critical issue. Many 2D magnetic materials, such as CrI$_3$ and Cr$_2$Ge$_2$Te$_6$, have Curie temperatures far below ambient conditions, limiting their practical applicability\cite{Lin2022, Hao2022}. Overcoming this challenge will require materials engineering strategies, such as strain modulation and chemical doping, to enhance thermal stability\cite{Ansari2021}. Additionally, the development of heterostructures that combine 2D magnets with non-magnetic layers has shown promise in enhancing coercivity and extending Curie temperatures beyond room temperature\cite{Dieny2020}.

Interfacial engineering also presents a major hurdle. Efficiently injecting, transporting, and manipulating spins in 2D materials necessitates pristine interfaces with minimal scattering and high polarization efficiency\cite{Dieny2020}. Techniques such as encapsulation with h-BN or advanced passivation strategies are crucial to mitigating interfacial degradation and improving device performance\cite{Hao2022}.

Despite all challenges, some emerging trends in spintronics are shaping the future of energy-efficient and high-performance computing. Next-generation 2D materials, such as magnetic vdW materials and topological insulators, offer exceptional properties like high spin polarization, low-dimensional confinement, and compatibility with existing semiconductor technologies. Room-temperature quantum spintronics is another transformative area, enabling the manipulation of spin states without requiring extreme cooling, which is crucial for practical applications in neuromorphic and quantum computing. Additionally, AI-enabled material discovery accelerates the identification and design of novel materials with optimized magnetic, electronic, and thermal properties, facilitating the development of more efficient spintronic devices. The 3D integration of 2D spintronic networks further enhances scalability and performance by enabling dense, multi-layer architectures that combine memory and logic functions, overcoming the limitations of traditional planar designs. Together, these advancements promise to revolutionize the field by improving energy efficiency, enhancing computational capabilities, and enabling new paradigms for information processing.

In conclusion, addressing the challenges and taking advantage of emerging trends will be key to advancing 2D spintronic devices. Future research directions, particularly focusing on hybrid systems and 3D integration, will unlock the full potential of these materials for transformative neuromorphic computing applications. Moreover, exploring unconventional approaches such as multi-functional 2D magnets and quantum spin liquids may open new frontiers in neuromorphic computing. Additionally, the key questions surrounding AI creativity-what it truly means for a machine to be creative, how AI can contribute to creative processes, and whether AI can genuinely replicate human-like innovation—remain critical. We now know that AI has the potential to generate novel ideas and even mimic artistic processes, but whether it can truly possess creativity in the human sense is still debated. This matters because understanding AI's role in creativity will shape how it integrates into industries, disrupts traditional work models, and transforms the creative sectors. The implications for the future of AI and work are profound, as AI could both complement and challenge human workers in creative fields. However, significant research gaps remain, particularly in terms of defining and measuring creativity in machines. Future directions should focus on advancing AI's capabilities using neuromorphic architecture, and there is a pressing need for interdisciplinary collaboration between AI researchers, artists, and philosophers to address these challenges and explore the full potential of AI creativity.
These insights highlight the transformative potential of 2D spintronics and electronics in shaping the future of energy-efficient, high-performance computing systems.

\section*{Author Contributions}
Y. Wu conceived the ideas and supervised the project. D. Z. Plummer led the review paper writing and discussions, with contributions from E. D'Alessandro, A. Burrowes, J. Fleischer and A. M. Heard. 

\section*{Funding}
Support from UF Gatorade award, Research Opportunity Seed Fund, and National Science Foundation are kindly acknowledged.

\section*{Acknowledgments}
Support from UF Gatorade award and Research Opportunity Seed Fund are kindly
acknowledged. This material is also based upon work supported by the National Science Foundation under Grant No. 2441051. 

\section*{Conflicts of Interest}
The authors declare no conflicts of interest. The funders had no role in the design of the study; in the collection, analyses, or interpretation of data; in the writing of the manuscript; or in the decision to publish the results.

\section*{References}
\bibliographystyle{unsrt}
\bibliography{ref.bib}
\end{document}